%% file: nova_nue_sa.tex
\documentclass[aps,prl,twocolumn,showpacs,superscriptaddress,preprintnumbers]{revtex4}  
\usepackage{graphicx}  
\usepackage{dcolumn}   
\usepackage{bm}        
\usepackage{amssymb}   
\usepackage{amsmath}   
\usepackage{multirow}
\usepackage{color}
\usepackage{units}
\usepackage{xspace}

\newcommand{\etal}{{\it et al.}\xspace}

\newcommand{\nova}{NOvA\xspace}
\newcommand{\numu}{\ensuremath{\nu_{\mu}}\xspace}
\newcommand{\nue}{\ensuremath{\nu_{e}}\xspace}
\newcommand{\numubar}{\ensuremath{\bar{\nu}_{\mu}}\xspace}
\newcommand{\nuebar}{\ensuremath{\bar{\nu}_{e}}\xspace}

\newcommand{\nueCC}{\ensuremath{\nu_e\,{\rm CC}}\xspace}
\newcommand{\numuCC}{\ensuremath{\nu_{\mu}\,{\rm CC}}\xspace}
\newcommand{\nutau}{\ensuremath{\nu_{\tau}}\xspace}
\newcommand{\dmsq}[1]{\ensuremath{\Delta m^2_{ #1 }}\xspace}

\newcommand{\dcp}{\ensuremath{\delta_{CP}}\xspace}

\newcommand{\anue}{\mbox{$\overline{\nu}_{e}$}\xspace}          
\newcommand{\anumu}{\mbox{$\overline{\nu}_{\mu}$}\xspace}       
\newcommand{\piz}{\mbox{$\pi^{0}$}\xspace}
\newcommand{\nutauCC}{\ensuremath{\nu_{\tau}\,{\rm CC}}\xspace}
\newcommand{\POT}{POT}

\newcommand{\ihlothreesigallowedrange}{\ensuremath{0.97\pi<\dcp<1.94\pi}\xspace}

\begin{document}

\title{Constraints on Oscillation Parameters from \nue Appearance and \numu Disappearance in NOvA}

\input novanue2016.tex

\date{\today}

\preprint{FERMILAB-PUB-17-065-ND}

\begin{abstract}
  Results are reported from an improved measurement of $\numu\to\nue$ transitions
  by the \nova experiment. Using an exposure equivalent to $6.05\times10^{20}$
  protons-on-target 33 $\nu_e$ candidates were observed with a background of 
  $8.2\pm0.8$ (syst.). Combined with the latest \nova $\nu_\mu$ disappearance
  data and external constraints from reactor experiments on $\sin^22\theta_{13}$, the hypothesis of inverted mass hierarchy with $\theta_{23}$  in the lower octant is disfavored at greater than $93\%$ C.L. for all values of \dcp.
\end{abstract}

\maketitle

This Letter reports updated results on the rate of $\numu\to\nue$ transitions in the \nova experiment~\cite{ref:novaFAnue} and constraints on oscillation parameters from the first combined fit of \nue appearance and \numu disappearance data.
The measurement, also probed by MINOS ~\cite{ref:minosnumu} and T2K~\cite{ref:t2knumu} experiments, is sensitive to three unknowns in neutrino physics: the octant of $\theta_{23}$ (whether $\theta_{23}$ is less than, equal to, or greater than $\pi/4$), the neutrino mass hierarchy, and the amount of CP violation in the lepton sector.
At the baseline and neutrino energy range of the NOvA experiment the probability for $\nu_\mu$ to oscillate to $\nu_e$ is primarily proportional to the combination $\sin^2\theta_{23}\sin^22\theta_{13}$.
The disappearance of muon neutrinos is sensitive to the mixing angle $\theta_{23}$ which is relatively weakly constrained to be
near-maximal ($\sin^2\theta_{23}\approx0.5$)~\cite{ref:minosnumu, ref:t2knumu, ref:novaFAnumu}.
Reactor neutrino measurements tightly constrain $\sin^22\theta_{13}$ at $0.085\pm0.005$ \cite{ref:dayabay, ref:reno, ref:doublechooz}.
The coherent forward scattering of the neutrino beam with electrons in the Earth enhances the electron neutrino appearance probability in the case of normal mass hierarchy (NH), where $\Delta m^2_{32}>0$, and suppresses it for inverted mass hierarchy (IH), where $\Delta m^2_{32}<0$.
The possible violation of CP symmetry in the lepton sector is parameterized by  \dcp.
CP-conserving oscillations occur if $\dcp=0$ or $\pi$, while $\nu_e$ appearance is enhanced around $\dcp=3\pi/2$, and suppressed around $\dcp=\pi/2$.
At NOvA's energy and baseline the impact of these three factors on the \nue appearance probability are of similar magnitudes, which can lead
to degeneracies between them, particularly when analyzing oscillations in neutrinos alone.
For antineutrinos, the mass hierarchy and CP phase have the opposite effect on the oscillation probability, while increasing values of $\sin^2\theta_{23}$ increase the appearance probabilities for \nue and $\bar\nu_e$ alike.

NOvA \cite{ref:nova} observes neutrinos produced in Fermilab's NuMI \cite{ref:NuMI} beamline in two detectors.
The Far Detector (FD) is located on the surface, \unit[14.6]{mrad} off the central beam axis, \unit[810]{km} from the neutrino parent production source.
The Near Detector (ND) is located \unit[100]{m} underground, \unit[1]{km} from the source and measures the neutrino beam spectrum before oscillations occur.
It is positioned to maximize the overlap between the neutrino energy spectra observed at the two detectors.
At these locations, the beam is peaked around \unit[2]{GeV} with neutrino energies mainly in the 1 to \unit[3]{GeV} range.
According to simulations, the neutrino beam at the ND is predominantly \numu, with 1.8\% \numubar and 0.7\% \nue + \nuebar components for neutrino energies between 1 and \unit[3]{GeV}.

The two functionally equivalent detectors \cite{ref:nova, ref:detector, ref:novaFAnumu, ref:novaFAnue} are constructed from planes of extruded PVC cells \cite{ref:extrusions}. 
The cells have a rectangular cross section measuring \unit[3.9]{cm} by \unit[6.6]{cm} and are \unit[15.5]{m} (\unit[3.9]{m}) long in the FD (ND).
Planes alternate the long cell dimension between vertical and horizontal orientations perpendicular to the beam.
Each cell is filled with liquid scintillator \cite{ref:scint}.
Light is collected by a loop of wavelength-shifting fiber inside the cell.
The fiber ends terminate on a single pixel of an avalanche photodiode (APD) \cite{ref:apd}.
The FD (ND) has a total active mass of \unit[14]{kt} (\unit[193]{t}).
In the fiducial region, the detectors are 62\% scintillator by mass.

The data analyzed were collected between February 6, 2014 and May 2, 2016.
The exposure is equivalent to $6.05\times 10^{20}$ protons-on-target (\POT) collected in the full detector and corresponds to more than double the exposure used in previous results \cite{ref:novaFAnumu, ref:novaFAnue}.
The fiducial mass for the full detector is \unit[10.3]{kt}.
The average neutrino beam power increased from \unit[250]{kW} to \unit[560]{kW}  during the data-taking period.

Measuring electron-neutrino appearance requires identification of charged-current (CC) interactions of \nue and understanding the various backgrounds that are also selected at the FD.
The signature of \nue CC interactions in the \nova detectors is an electromagnetic shower plus any associated hadronic recoil energy.
The largest background arises from Neutral Current (NC) interactions of beam neutrinos that produce \piz which decay to photons that mimic the signature of an electron.
The intrinsic \nue component of the NuMI beam represents an irreducible background to this search. 
Charged current interactions of \numu with a short muon track and a hadronic shower with some electromagnetic activity comprise a smaller background.
Other small backgrounds include cosmic ray induced events, particularly where a photon or a neutron enters from the sides of the detector and charged-current interactions of \nutau, which mostly occur above 3 GeV.

For this analysis a new \nueCC classifier was developed to select a signal sample with improved purity and efficiency.
The Convolutional Visual Network (CVN)~\cite{cvnpaper} is a convolutional neural network and was designed using deep learning techniques from the field of computer vision~\cite{szegedy2014googlenet,hinton1986}.
Recorded hits in the detectors are formed into clusters by grouping hits in time and space to isolate individual interactions \cite{ref:michaelthesis, ref:dbscan}.
The CVN classifier takes the hits from these clusters, without any further reconstruction, as input and applies a series of trained linear operations to extract complex, abstract classifying features from the image.
A multilayer perceptron~\cite{ref:mlp1,ref:mlp2} at the end of the network uses these features to create the classifier output.
Training is conducted using a mixture of simulated FD \numuCC, \nueCC, \nutauCC, and NC events as well as a sample of FD cosmic data.

The \nova simulation chain uses FLUKA~\cite{ref:fluka1}, GEANT4~\cite{ref:geant1}, FLUGG~\cite{ref:flugg}, GENIE~\cite{ref:genie} and a custom detector simulation~\cite{ref:sim} to model neutrino production in the beamline and subsequent interaction in the detector.
Neutrino scattering off substructure in the nucleus is added to the simulation using an empirical model of multinucleon excitations and long range correlations \cite{ref:mec1,ref:mec2,ref:mec3,ref:mec4}.
The implementation of this model in the \nova simulation is tuned to match an observed excess of events in data over simulation in bins of reconstructed three-momentum transfer~\cite{ref:numuSA}.
Additionally, the rate of non-resonant single pion production in charged-current interactions is effectively reduced by 50\%, motivated by a recent reanalysis of \numu-deuterium pion-production data \cite{ref:nonres1pi1,ref:nonres1pi2}.

For the purpose of energy reconstruction and event containment, the event cluster is further reconstructed to determine particle paths.
A Hough transform is applied  to  the  cluster  to  identify global features, characterized as Hough lines\cite{ref:hough}.
The intersections of these lines seed an algorithm to produce a three dimensional vertex for the cluster \cite{ref:earms}.
 In both the horizontal and vertical detector views hits are grouped into prongs radiating from the vertex \cite{ref:fuzzyk,ref:evanthesis}.
 Prongs are then matched between the views based on energy deposition characteristics.

The energy responses of the detectors are calibrated using minimum ionizing energy deposits in a region 1 to \unit[2]{meters} from the end of tracks corresponding to stopping cosmic ray muons.
To reconstruct the electron neutrino candidate energy, the prong with the most calorimetric energy is assumed to be an electromagnetic shower caused by the outgoing electron.
The remaining energy deposits in the event are attributed to the hadronic recoil system.
The reconstructed \nue energy is taken as a quadratic function of the electromagnetic and hadronic calorimetric responses.
The function is a parameterization of the simulated true electron neutrino energy in relation to these quantities, and yields an energy resolution of $\sim$7\% in both detectors.

To suppress the cosmic ray induced background in the FD, selected events are required to be in a \unit[12]{$\mu $s} window centered on the \unit[10]{$\mu $s} beam spill.
A large fraction of cosmic events deposit energy close to the detector edges and are removed due to containment requirements.
Requiring a small reconstructed transverse momentum fraction with respect to the beam direction rejects cosmic events with angles too steep to be consistent with a NuMI beam event.
The cosmic background rejection criteria are tuned using neutrino beam simulation and a large sample of cosmic data recorded asynchronously with the neutrino beam.

The maximum of the \nue appearance signal is expected just below the peak neutrino energy at \nova.
Restricting the energy range of selected events to \unit[1-3]{GeV} removes a large fraction of the NC and cosmic backgrounds which are predominately of lower reconstructed energy, and intrinsic \nueCC events which dominate at higher energies.
We similarly constrain the length of the longest track and number of hits in an event to remove  clear muon tracks or poorly reconstructed events.
Other than containment requirements, the \nue CC selection criteria in the ND are very similar to those in the FD.

The selection criteria are chosen to maximize the figure of merit defined as $\frac{S}{\sqrt{S+B}}$, where {\it S} and {\it B} are the number of signal and background events, respectively.
The final \nue selection criteria select contained appearance signal with $73.5\%$ efficiency and $75.5\%$ purity, representing a gain in sensitivity of $30\%$  compared to the \nue classifiers used in the previously reported results \cite{ref:novaFAnue}.
These criteria also reject $97.6\%$ of the NC and $99.0\%$ of the \numu CC beam backgrounds.
The cosmic ray backgrounds are suppressed by  seven orders of magnitude, and only $0.53\pm 0.14$ cosmic events are estimated to be selected in the final \nue appearance sample based on the performance of \nue selection criteria on cosmic data.
Of the beam backgrounds that pass all \nue selection, 91\% contain some form of energetic electromagnetic shower.
To further improve the statistical power of this analysis, events selected in the FD are split into three \nue classifier bins, containing signal \nue CC events with low, medium and high purity.
The analysis is performed in four energy bins between 1 and \unit[3]{GeV} for each of the three classifier bins.

\begin{figure}
  \includegraphics[width=\linewidth]{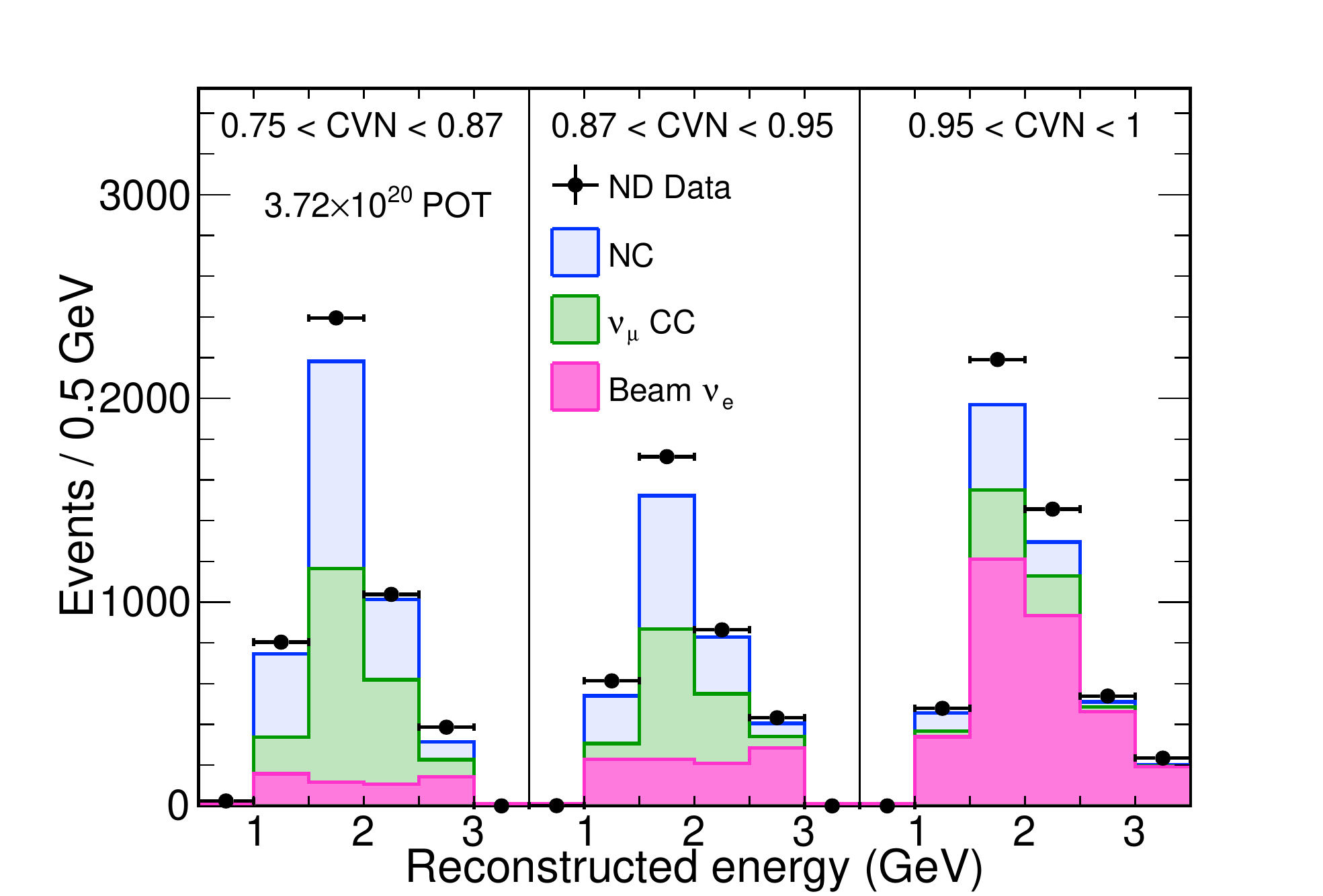}
  \caption{ Reconstructed energy of events selected in the ND data and simulation by the \nue CC selection criteria  in the three \nue classifier (CVN) bins. The left-most panel is the lowest purity classifier bin, while the right-most has the highest purity. 
}
  \label{fig:nd_spectrum}
\end{figure}

The ND has negligible \nue appearance signal, and is used to estimate the beam neutrino induced background rates to the appearance measurement.
According to simulation, the kinematics of the events that pass the \nue CC selection criteria in the ND are representative of and adequately cover those selected in the FD.
Figure \ref{fig:nd_spectrum} shows that there is an overall $\sim$10\% excess of data over simulation in the \nue CC selected events in the ND.
Since the NC, \numu CC and beam \nue CC background components are affected differently by oscillations, the total background selected in the ND data is broken down into these components which are then used to estimate the corresponding components in the FD.

Both the \numu and intrinsic \nue components of the beam peak arise primarily from pions decaying through the process ($\pi^+ \rightarrow \mu^+ + \nu_\mu$), as well as the subsequent muon decay ($\mu^+ \rightarrow e^+ + \numubar + \nu_e$).
At higher energies they originate from kaon decays.
The pion and kaon hadron yields can be derived from the low and high-energy \numuCC rate in the ND data and are used to correct the \nue CC rate in the simulation.
Pion yields are adjusted in bins of transverse and longitudinal pion momentum, while the kaon yield is simply scaled.
From this method, it is inferred that the kaon yield is higher by 17\% and the pion yield lower by 3\% than predicted by the simulation. 
This results in an overall 1\% increase in the estimated intrinsic \nueCC background rate in the 1 to \unit[3]{ GeV} range in the ND.

Some of the \numuCC interactions that are a background to the \nueCC selection have a muon hidden in the shower associated with the hadronic recoil.
In these events, the time-delayed electron from muon decay (Michel electron) may often be found.
The hadronic recoil system also produces this signature due to the presence of charged pions that decay to muons.
However, on average, \numuCC interactions have one more Michel electron than \nueCC and NC interactions.
The \numuCC and NC background components are varied in each bin of energy and \nue classifier to obtain the best match to the distribution of the number of Michel electron candidates in data.
The intrinsic \nueCC background component is held fixed at the value obtained from the pion and kaon yield analysis.
This method leads to an integrated increase of 17.7\% and 10.4\% in the \numuCC and NC background rates relative to those predicted by the ND simulation.
These corrections derived from the ND data account for the 10\% discrepancy with simulation and are applied to the background spectra in the FD simulation in the analysis bins.
The spectra are then weighted by the appropriate 3-flavor oscillation probability to obtain the final estimates of the beam backgrounds in the FD.
After applying these data-driven constraints, the predicted background composition in the FD for this analysis is 45.3\% NC, 38\% intrinsic \nue CC, 8.4\% \numu CC, 1.8\% \nutauCC, and 6.5\% cosmic events.

\begin{figure}
  \includegraphics[width=\linewidth]{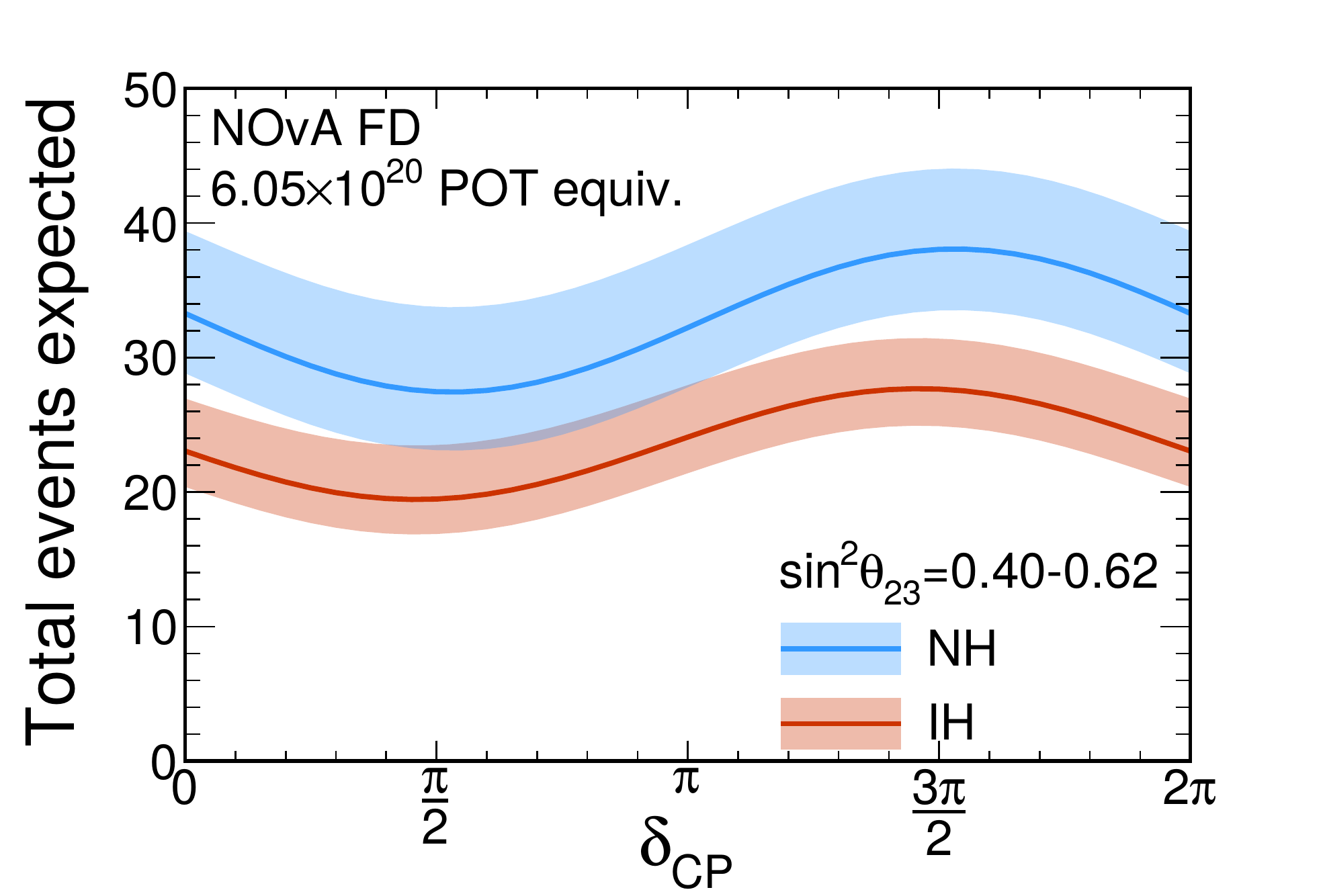}
  \caption{Total number of selected $\nu_e$ candidate events expected at the FD. The blue represents Normal Hierarchy (NH) and the orange Inverted Hierarchy (IH). The bands correspond to the range $\sin^2\theta_{23}=0.40$ (lower edge) to $0.62$ (upper edge), with the solid line marking maximal mixing. The $x$-axis gives the value of the CP phase, while all other parameters are held fixed at the best fit values found by NOvA's latest analysis of $\nu_\mu$ disappearance \cite{ref:numuSA}.
  }
  \label{fig:monoprob}
\end{figure}

The \nue appearance signal expected in the FD is also constrained by the observed neutrino beam spectrum in the ND.
A sample of \numu candidates are selected in the ND data using the latest \numu
selection criteria as described in \cite{ref:numuSA}, and the underlying true energy spectrum is derived from a reconstructed to true energy migration matrix.
The spectrum of true \nueCC signal events selected in the FD simulation is corrected by the ratio of the \numuCC true energy spectrum derived from ND data to the simulated \numuCC spectrum.
The adjusted FD signal spectrum is weighted by the \nue appearance probability and mapped back to the reconstructed energy spectrum for the final estimate of \nue appearance signal.
This extrapolation is carried out for the energy spectra in all three \nue classifier bins.
Figure \ref{fig:monoprob} shows the variation in the number of FD events predicted as a function of the assumed oscillation parameters.

The ND data are also used to verify the simulated \nue CC selection efficiency.
For events that pass the \numu CC selection criteria in ND data and simulation, the energy deposits along the reconstructed track of the candidate muon are removed \cite{ref:kanikathesis}.
An electron with the same energy and direction is simulated in its place to construct \nueCC-like interactions in both data and simulation.
The event is reconstructed again with the electron shower embedded in it and the \nue selection cuts are applied.
The efficiency of the \nue CC selection criteria in the ND between data and simulation for identifying neutrino events with inserted electrons matches to within 1\%.

\begin{figure}
  \includegraphics[width=\linewidth]{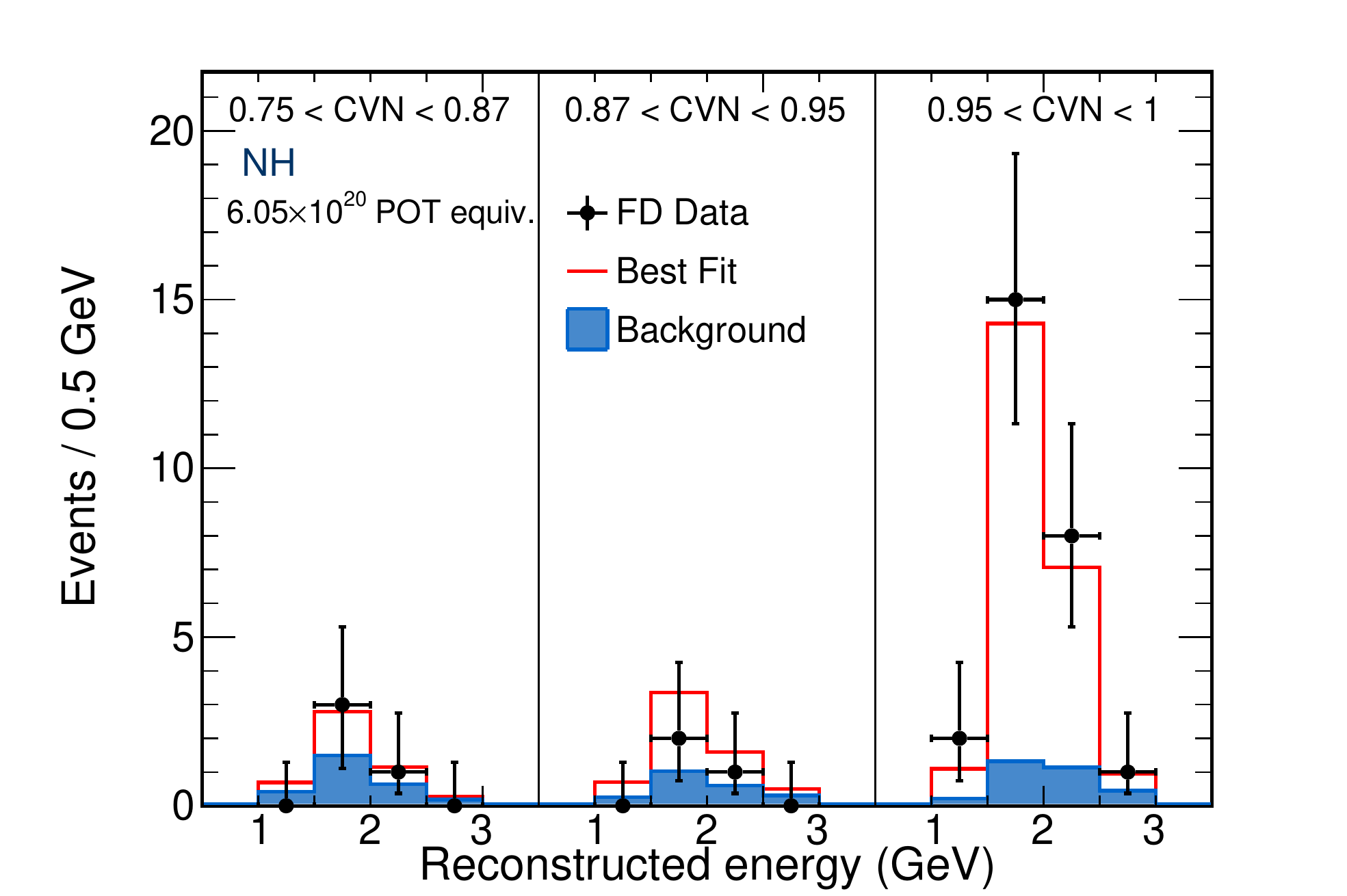}
  \caption{Reconstructed energy of selected FD events in three bins of the CVN
    classifier variable. Black points show the data, the red line shows the predicted spectrum at the best fit point in Normal Hierarchy (NH),
  with the blue area showing the total expected
    background.
  }
  \label{fig:fd_spectrum}
\end{figure}

Systematic uncertainties are evaluated by reweighting or generating new simulated event samples modified to account for each uncertainty in the ND and FD.
The full analysis, including background component estimation in the ND data and extrapolation to FD, is performed with these systematically shifted simulation samples to predict the altered signal and background spectra at the FD.
Calibration and normalization are the leading sources of systematic uncertainty for background and signal, respectively.
Other sources of systematic uncertainty considered include neutrino flux, modeling of neutrino interactions and detector response.
The overall effect of the uncertainties summed in quadrature on the total event count is 5.0\% (10.5\%) on the signal (background).
The statistical uncertainties of 20.1\% (34.9\%) on the signal (background) therefore dominate.

\begin{figure}
  \includegraphics[width=\linewidth,trim=0 55 0 0,clip]{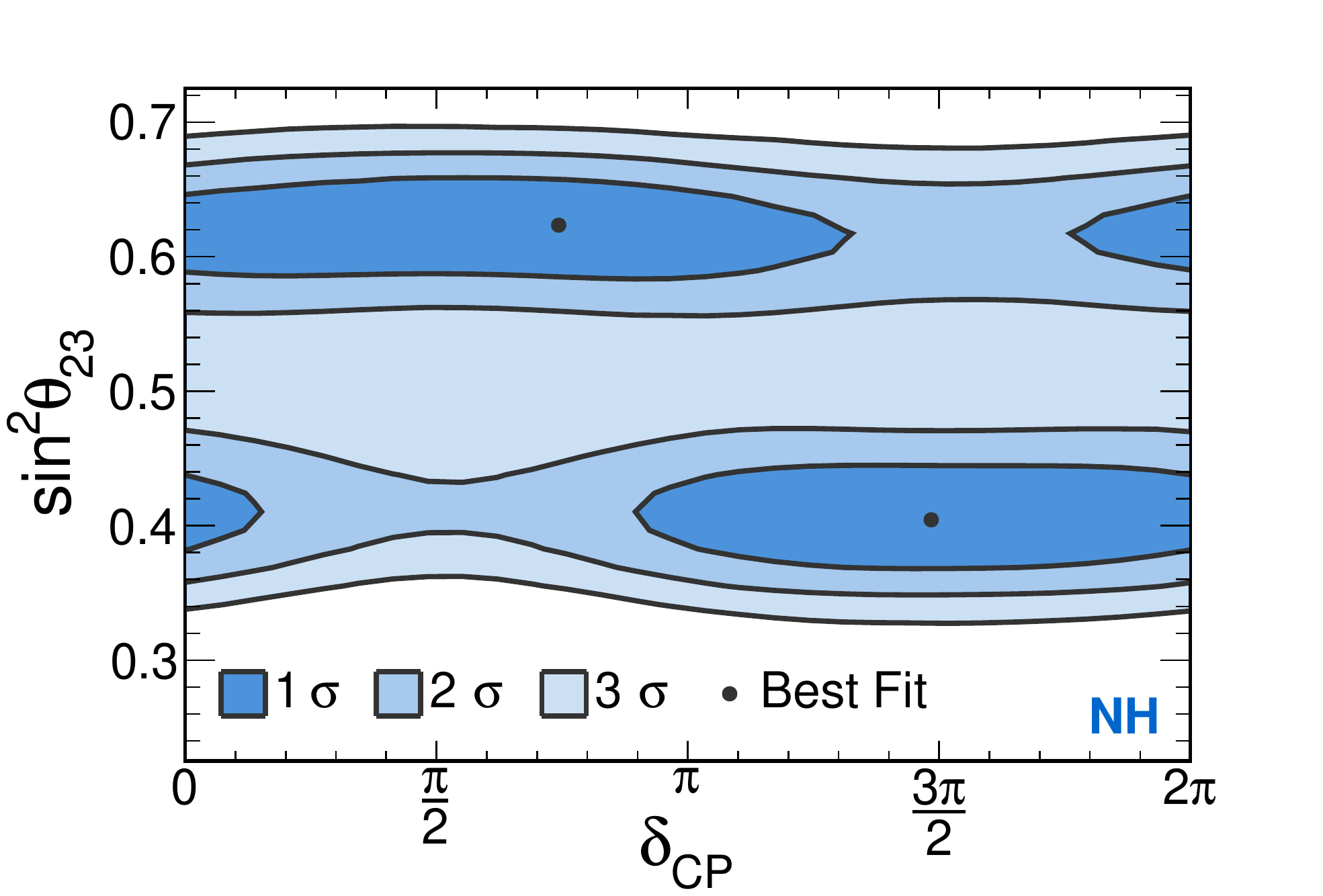}\\
  \vspace{-.5em}
  \includegraphics[width=\linewidth,trim=0 0 0 32,clip]{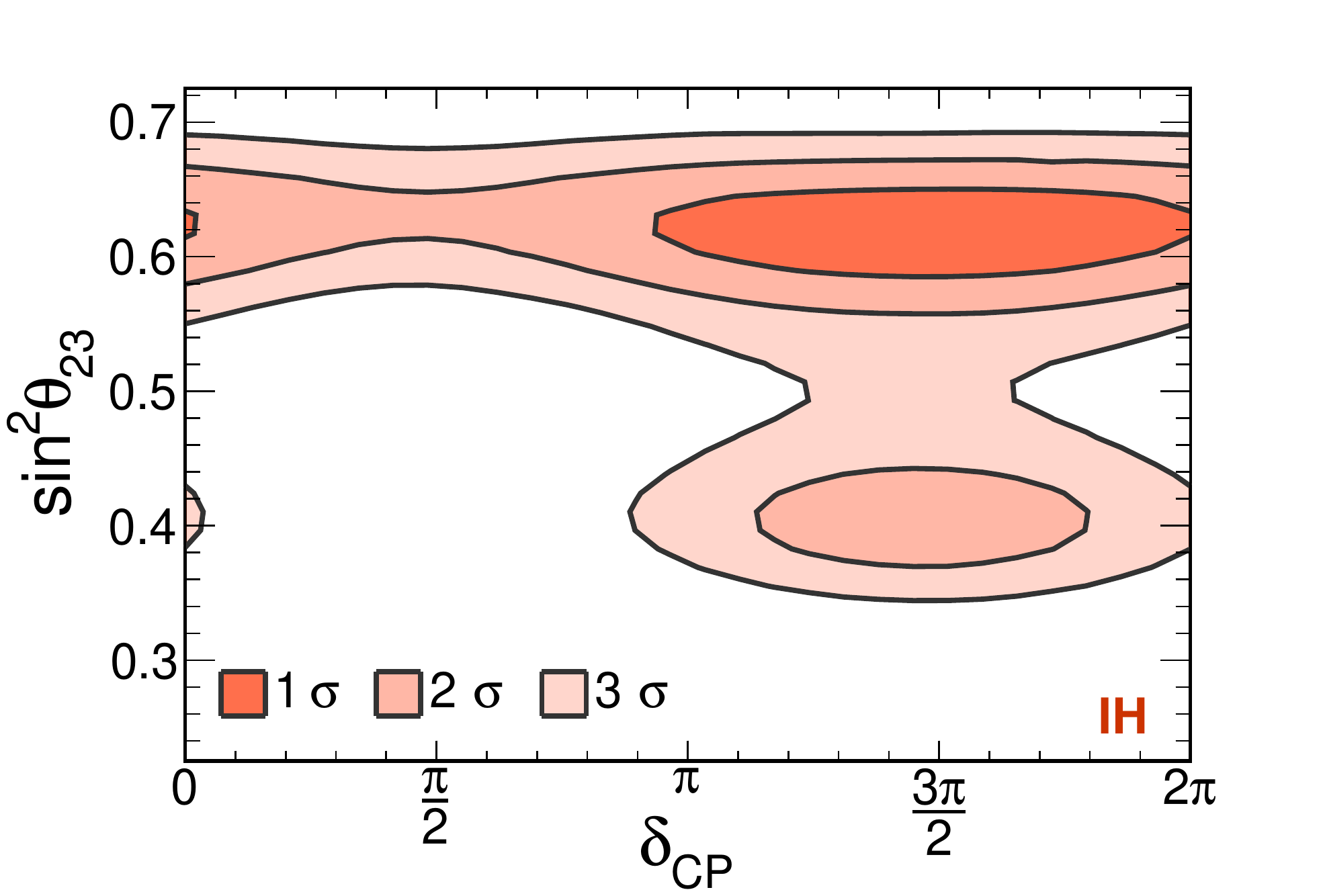}

  \caption{Regions of \dcp vs. $\sin^2\theta_{23}$ parameter space consistent
    with the observed spectrum of $\nu_e$ candidates and the \numu
    disappearance data \cite{ref:numuSA}. The top panel corresponds to normal
    mass hierarchy ($\dmsq{32}>0$) and the bottom panel to inverted hierarchy
    ($\dmsq{32}<0$). The color intensity indicates the confidence level at which
    particular parameter combinations are allowed.}
  \label{fig:wave}
\end{figure}

After the event selection criteria and analysis procedures were finalized, inspection of the FD data revealed 33 $\nu_e$ candidates, of which $8.2\pm 0.8$ (syst.) events are predicted to be background \footnote{The backgrounds are computed at the best fit oscillation parameters: $\sin^2\theta_{23} = 0.40,\ \sin^22\theta_{13} = 0.085, \Delta m_{32}^2 = 2.67 \times 10^{-3}\ {\rm eV}^2 \rm{and}\ \dcp = 1.48 \pi$. The matter density, computed for the average depth of the NuMI beam in the earth crust for the NOvA baseline of \unit[810]{km} using the CRUST2.0 \cite{ref:crust} model, is \unit[2.84]{g/cm$^3$.} }.
Figure \ref{fig:fd_spectrum} shows a comparison of the event distribution with the expectations at the best fit point as a function of the classifier variable and reconstructed neutrino energy.

To extract oscillation parameters, the \nue CC energy spectrum in bins of event classifier is fit simultaneously with the FD \numu CC energy spectrum~\cite{ref:numuSA}.
The \nova \numu disappearance result constrains $\sin^2\theta_{23}$ around degenerate best fit points of 0.404 and 0.624.
The likelihood between the observed spectra and the Poisson expectation in each bin is computed as a function of the oscillation parameters $|\Delta m^2_{32}|$, $\theta_{23}$, $\theta_{13}$, \dcp, and the mass hierarchy.
Each source of systematic uncertainty is incorporated into the fit as a nuisance parameter, which varies the predicted FD spectrum according to the shifts determined from systematically shifted samples.
Where systematic uncertainties are common between the two data sets, the nuisance parameters associated with the effect are correlated appropriately.
Gaussian penalty terms are applied to represent the estimates of the $1\sigma$ ranges of these parameters, and the knowledge of $\sin^22\theta_{13}=0.085\pm0.005$ from reactor experiments \cite{ref:PDG}.

Figure~\ref{fig:wave} shows the regions of ($\sin^2\theta_{23}$, \dcp) space allowed at various confidence levels.
The likelihood surface is profiled over the parameters $|\Delta m^2_{32}|$ and $\theta_{13}$ while the solar parameters $\Delta m^2_{21}$ and $\theta_{12}$ are held fixed.
The significances are derived using the Feldman-Cousins unified approach \cite{ref:fc} to account for the statistical effects of low event count and physical boundaries.

\begin{figure}
  \includegraphics[width=\linewidth]{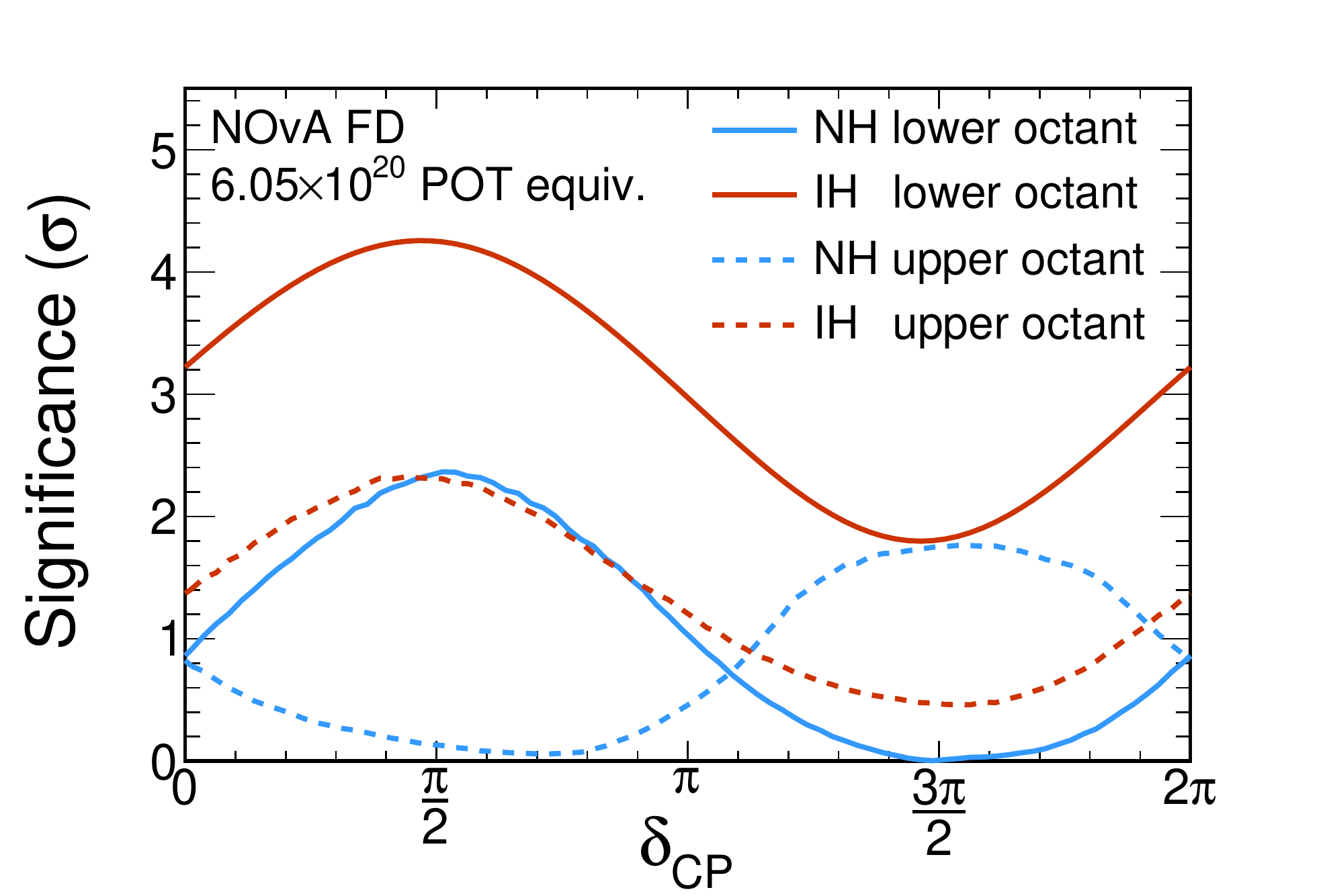}
  \caption{Feldman-Cousins corrected significance at which each value of \dcp is disfavored for each of
    the four possible combinations of mass hierarchy: normal (blue) or
    inverted (red), and $\theta_{23}$ octant: lower (solid) or upper (dashed),
    by the combination of $\nu_e$ appearance and NOvA's latest \numu
    disappearance measurement \cite{ref:numuSA}.}
  \label{fig:slice}
\end{figure}

Figure~\ref{fig:slice} shows the significance at which values of \dcp are disfavored for each hierarchy and octant combination.
The value of $\sin^2\theta_{23}$ is profiled within the specified octant.
There are two degenerate best fit points, both in the normal hierarchy, $\sin^2\theta_{23} = 0.404$, $\dcp = 1.48\pi$ and $\sin^2\theta_{23} = 0.623$, $\dcp = 0.74\pi$.
The inverted hierarchy predicts fewer events than are observed for all values of \dcp and both octants.
The best-fit point in the inverted hierarchy occurs near $\dcp = 3\pi/2$ and is 0.46 $\sigma$ from the global best-fit points.
The inverted mass hierarchy in the lower octant is disfavored at greater than 93\% C.L. for all values of \dcp, and excluded at greater than $3\sigma$ significance outside the range \ihlothreesigallowedrange.
The T2K collaboration has recently published results based on their observation of \numu(\anumu) disappearance and \nue(\anue) appearance ~\cite{ref:t2k2017}.
While their data favor a near-maximal value of $\theta_{23}$, they disfavor CP conservation at 90\% C.L., with a weak preference for normal mass hierarchy.
These observations are broadly consistent with the NOvA result.

To conclude, in the first combined fit of the \nova \nue appearance and \numu disappearance data, the inverted mass hierarchy with $\theta_{23}$  in the lower octant is disfavored at greater than $93\%$ C.L. for all values of \dcp.
Future data-taking in antineutrino mode, where the impact of the mass hierarchy and CP phase are reversed with respect to their effect on neutrinos, will help resolve the remaining degeneracies in the parameters.

This work was supported by the US Department of Energy; the US National Science Foundation; the Department of Science and Technology, India; the European Research Council; the MSMT CR, GA UK, Czech Republic; the RAS, RMES, and RFBR, Russia; CNPq and FAPEG, Brazil; and the State and University of Minnesota. We are grateful for the contributions of the staffs at the University of Minnesota module assembly facility and Ash River Laboratory, Argonne National Laboratory, and Fermilab. Fermilab is operated by Fermi Research Alliance, LLC under Contract No.~De-AC02-07CH11359 with the US DOE.

\end{document}

%% file: novanue2016.tex
\newcommand{\ANL}{Argonne National Laboratory, Argonne, Illinois 60439,
USA}
\newcommand{\IOP}{Institute of Physics, The Czech Academy of Sciences,
182 21 Prague, Czech Republic}
\newcommand{\BHU}{Department of Physics, Institute of Science, Banaras
Hindu University, Varanasi, 221 005, India}
\newcommand{\UCLA}{Physics and Astronomy Department, UCLA, Box 951547, Los
Angeles, California 90095-1547, USA}
\newcommand{\Caltech}{California Institute of
Technology, Pasadena, California 91125, USA}
\newcommand{\Cochin}{Department of Physics, Cochin University
of Science and Technology, Kochi 682 022, India}
\newcommand{\Charles}
{Charles University, Faculty of Mathematics and Physics,
 Institute of Particle and Nuclear Physics, Prague, Czech Republic}
\newcommand{\Cincinnati}{Department of Physics, University of Cincinnati,
Cincinnati, Ohio 45221, USA}
\newcommand{\CSU}{Department of Physics, Colorado
State University, Fort Collins, CO 80523-1875, USA}
\newcommand{\CTU}{Czech Technical University in Prague,
Brehova 7, 115 19 Prague 1, Czech Republic}
\newcommand{\Dallas}{Physics Department, University of Texas at Dallas,
800 W. Campbell Rd. Richardson, Texas 75083-0688, USA}
\newcommand{\Delhi}{Department of Physics and Astrophysics, University of
Delhi, Delhi 110007, India}
\newcommand{\JINR}{Joint Institute for Nuclear Research,
Dubna, Moscow region 141980, Russia}
\newcommand{\FNAL}{Fermi National Accelerator Laboratory, Batavia,
Illinois 60510, USA}
\newcommand{\UFG}{Instituto de F\'{i}sica, Universidade Federal de
Goi\'{a}s, Goi\^{a}nia, Goi\'{a}s, 74690-900, Brazil}
\newcommand{\Guwahati}{Department of Physics, IIT Guwahati, Guwahati, 781
039, India}
\newcommand{\Harvard}{Department of Physics, Harvard University,
Cambridge, Massachusetts 02138, USA}
\newcommand{\IHyderabad}{Department of Physics, IIT Hyderabad, Hyderabad,
502 205, India}
\newcommand{\Hyderabad}{School of Physics, University of Hyderabad,
Hyderabad, 500 046, India}
\newcommand{\Indiana}{Indiana University, Bloomington, Indiana 47405,
USA}
\newcommand{\INR}{Inst. for Nuclear Research of Russia, Academy of
Sciences 7a, 60th October Anniversary prospect, Moscow 117312, Russia}
\newcommand{\Iowa}{Department of Physics and Astronomy, Iowa State
University, Ames, Iowa 50011, USA}
\newcommand{\Irvine}{Department of Physics and Astronomy,
University of California at Irvine, Irvine, California 92697, USA}
\newcommand{\Jammu}{Department of Physics and Electronics, University of
Jammu, Jammu Tawi, 180 006, Jammu and Kashmir, India}
\newcommand{\Lebedev}{Nuclear Physics Department, Lebedev Physical
Institute, Leninsky Prospect 53, 119991 Moscow, Russia}
\newcommand{\MSU}{Department of Physics and Astronomy, Michigan State
University, East Lansing, Michigan 48824, USA}
\newcommand{\Duluth}{Department of Physics and Astronomy,
University of Minnesota Duluth, Duluth, Minnesota 55812, USA}
\newcommand{\Minnesota}{School of Physics and Astronomy, University of
Minnesota Twin Cities, Minneapolis, Minnesota 55455, USA}
\newcommand{\Oxford}{Subdepartment of Particle Physics,
University of Oxford, Oxford OX1 3RH, United Kingdom}
\newcommand{\Panjab}{Department of Physics, Panjab University,
Chandigarh, 106 014, India}
\newcommand{\RAL}{Rutherford Appleton Laboratory, Science and
Technology Facilities Council, Didcot, OX11 0QX, United Kingdom}
\newcommand{\SAlabama}{Department of Physics, University of
South Alabama, Mobile, Alabama 36688, USA}
\newcommand{\Carolina}{Department of Physics and Astronomy, University of
South Carolina, Columbia, South Carolina 29208, USA}
\newcommand{\SDakota}{South Dakota School of Mines and Technology, Rapid
City, South Dakota 57701, USA}
\newcommand{\SMU}{Department of Physics, Southern Methodist University,
Dallas, Texas 75275, USA}
\newcommand{\Stanford}{Department of Physics, Stanford University,
Stanford, California 94305, USA}
\newcommand{\Sussex}{Department of Physics and Astronomy, University of
Sussex, Falmer, Brighton BN1 9QH, United Kingdom}
\newcommand{\Tennessee}{Department of Physics and Astronomy,
University of Tennessee, Knoxville, Tennessee 37996, USA}
\newcommand{\Texas}{Department of Physics, University of Texas at Austin,
Austin, Texas 78712, USA}
\newcommand{\Tufts}{Department of Physics and Astronomy, Tufts University, Medford,
Massachusetts 02155, USA}
\newcommand{\UCL}{Physics and Astronomy Dept., University College London,
Gower Street, London WC1E 6BT, United Kingdom}
\newcommand{\Virginia}{Department of Physics, University of Virginia,
Charlottesville, Virginia 22904, USA}
\newcommand{\WSU}{Department of Mathematics, Statistics, and Physics,
Wichita State Univ.,
Wichita, Kansas 67206, USA}
\newcommand{\WandM}{Department of Physics, College of William \& Mary,
Williamsburg, Virginia 23187, USA}
\newcommand{\Winona}{Department of Physics, Winona State University, P.O.
Box 5838, Winona, Minnesota 55987, USA}
\newcommand{\Crookston}{Math, Science and Technology Department, University
of Minnesota -- Crookston, Crookston, Minnesota 56716, USA}
\newcommand{\deceased}{Deceased.}

\affiliation{\ANL}
\affiliation{\IOP}
\affiliation{\BHU}
\affiliation{\Caltech}
\affiliation{\Charles}
\affiliation{\Cincinnati}
\affiliation{\Cochin}
\affiliation{\CSU}
\affiliation{\CTU}
\affiliation{\Delhi}
\affiliation{\FNAL}
\affiliation{\UFG}
\affiliation{\Guwahati}
\affiliation{\Harvard}
\affiliation{\Hyderabad}
\affiliation{\IHyderabad}
\affiliation{\Indiana}
\affiliation{\INR}
\affiliation{\Iowa}
\affiliation{\Irvine}
\affiliation{\Jammu}
\affiliation{\JINR}
\affiliation{\Lebedev}
\affiliation{\MSU}
\affiliation{\Duluth}
\affiliation{\Minnesota}
\affiliation{\Panjab}
\affiliation{\SAlabama}
\affiliation{\Carolina}
\affiliation{\SDakota}
\affiliation{\SMU}
\affiliation{\Stanford}
\affiliation{\Sussex}
\affiliation{\Tennessee}
\affiliation{\Texas}
\affiliation{\Tufts}
\affiliation{\UCL}
\affiliation{\Virginia}
\affiliation{\WSU}
\affiliation{\WandM}
\affiliation{\Winona}

\author{P.~Adamson}
\affiliation{\FNAL}

\author{L.~Aliaga}
\affiliation{\FNAL}

\author{D.~Ambrose}
\affiliation{\Minnesota}

\author{N.~Anfimov}
\affiliation{\JINR}

\author{A.~Antoshkin}
\affiliation{\JINR}

\author{E.~Arrieta-Diaz}
\affiliation{\SMU}

\author{K.~Augsten}
\affiliation{\CTU}

\author{A.~Aurisano}
\affiliation{\Cincinnati}

\author{C.~Backhouse}
\affiliation{\Caltech}

\author{M.~Baird}
\affiliation{\Sussex}
\affiliation{\Indiana}

% ...

\collaboration{The NOvA Collaboration}
\noaffiliation